\documentclass[apj]{emulateapj}
\usepackage{epsfig}
\usepackage{epsfig,wrapfig,setspace}

\newcommand{\etal}{et al.}

\def\simlt{\mathrel{\hbox{\rlap{\hbox{\lower4pt\hbox{$\sim$}}}\hbox{$<$}}}}
\def\simgt{\mathrel{\hbox{\rlap{\hbox{\lower4pt\hbox{$\sim$}}}\hbox{$>$}}}}

\def\chandra{{\it Chandra}}
\def\xmm{{\it XMM-Newton}}

\def\psr{PSR~J1210$-$5226}
\def\pks{PKS~1209$-$51/52}

\def\simlt{\mathrel{\hbox{\rlap{\hbox{\lower4pt\hbox{$\sim$}}}\hbox{$<$}}}}
\def\simgt{\mathrel{\hbox{\rlap{\hbox{\lower4pt\hbox{$\sim$}}}\hbox{$>$}}}}
\def\cco{1E~1207.4$-$5209}
\def\kpsr{PSR~J1852$+$0040}
\def\ksnr{Kes~79}

\slugcomment{}

\shorttitle{First Glitch in \cco}
\shortauthors{Gotthelf \& Halpern}

\begin{document}

\title{
The First Glitch in a Central Compact Object Pulsar: \cco\
}

\author{E. V. Gotthelf and J. P. Halpern}

\affil{Columbia Astrophysics Laboratory, Columbia University,
New York, NY 10027-6601, USA; eric@astro.columbia.edu}

\begin{abstract}

  Since its discovery as a pulsar in 2000, the central compact object (CCO)
  \cco\ in the supernova remnant \pks\ had been a stable 0.424~s
  rotator with an extremely small spin-down rate and weak
  ($B_s\approx9\times10^{10}$~G) surface dipole magnetic field.
  In 2016 we observed a glitch from \cco\ of at least
  $\Delta f/f = (2.8\pm0.4)\times10^{-9}$, which is typical in size for
  the general pulsar population.  However, glitch activity
  is closely correlated with spin-down rate $\dot f$,
  and pulsars with $\dot f$ as small as that of \cco\ are
  never seen to glitch.  Unlike in glitches of ordinary pulsars, 
  there may have been a large increase in $\dot f$ as well.
  The thermal X-ray spectrum of \cco, with its unique cyclotron
  absorption lines that measure the surface magnetic field strength,
  did not show any measurable change after the glitch, which rules out a major
  disruption in the dipole field as a cause or result of the glitch.
  A leading theory of the origin and evolution of CCOs, involving prompt
  burial of the magnetic field by fall-back of supernova ejecta,
  might hold the explanation for the glitch.

\end{abstract}

\keywords{ISM: individual (\pks) --- pulsars: individual (\cco, \psr) --- stars: neutron}

\section {Introduction}

The group of $\approx 10$ central compact objects (CCOs) in supernova
remnants (SNRs) are defined by their steady surface thermal X-ray
emission, lack of surrounding pulsar wind nebula, and non-detection at any
other wavelength.  Of the eight well-established CCOs, three were found
to be pulsars \citep{zav00,got05,got09}.  Their spin-down properties
provide an estimate of surface dipole magnetic field strength via
$B_s=3.2\times10^{19}\sqrt{\dot f/f^3}$ of
$(2.8, 3.1, 9.8) \times 10^{10}$~G \citep{hal10,got13a},
two orders of magnitude less than those of  canonical young pulsars.
The homogeneous X-ray properties of the remaining CCOs that have not
yet been seen to pulse suggest that they have similar or even weaker
magnetic fields than the known CCO pulsars, or a more uniform surface
temperature, or a more aligned geometry.  The fact that CCOs are found
in SNRs in comparable numbers to other classes of young
neutron stars (NSs) implies that they must represent a significant 
fraction of NS births. 

\cco\ in the SNR \pks\ is the first CCO pulsar discovered \citep{zav00} and the most intensively
studied.  It is also the first isolated NS to display strong
absorption lines in its X-ray spectrum \citep{san02,mer02,big03,del04}.
The evenly spaced spectral features are widely accepted as the
electron cyclotron fundamental at $E_0=0.7$~keV and its harmonics
in a magnetic field of $B\approx8\times10^{10}$~G.
[More precisely, these features are due to quantum oscillations in the
free-free opacity \citep{sul10}]. 
This is the first measurement of the surface magnetic
field on a CCO by a direct technique that is independent of timing, and
the result is fully consistent with the magnetic field inferred from its
spin-down, $B_s=9.8\times10^{10}$~G.

In a series of papers \citep{got07,hal11,got13a,hal15} we have followed
the long-term timing properties of \cco\ which, until 2014, showed
steady spindown with no evidence of timing noise or glitches.
Glitches are sudden increases in spin frequency
that are thought to result from either ``starquakes,''
stress relief of the NS crust (e.g., \citealt{lin98}),
or the sudden unpinning and repinning of neutron superfluid
vortices in the inner crust \citep{alp84}.  Glitch activity among
pulsars is correlated mainly with frequency derivative,
such that pulsars with $|\dot f|$ as small as those of
CCOs have never been seen to glitch \citep{esp11,fue18}.
Nevertheless, interior properties of CCOs may be similar
to those of canonical young pulsars, which may cause them
to glitch \citep{ho15}.

We present new observations of \cco\ in 2016--2018
that detect a glitch for the first time in a CCO.
In Section 2, we describe the new X-ray timing observations.
Section 3 details the timing properties of the detected glitch, while 
Section~4 compares the pre- and post-glitch spectrum and flux
to search for any changes.  In Section~5, we compare the properties
of this glitch to the ensemble of measured NS glitches, and discuss
its implications for theories of the internal structure and evolution
of CCOs.  An alternative model of accretion-torque fluctuations is also
briefly considered.  Conclusions and suggestions for follow-up work are
described in Section~6.

\section{New X-ray Observations}
 
In the course of monitoring of \cco\ we found that
the phase of the pulsar measured in 2016 July no longer followed
the prediction of the prior ephemeris.
We obtained a new set of \xmm\ observations to confirm the glitch
and characterize its properties.  Six observations
in 2017 June--December were used
to bootstrap a new timing solution.  Then we began semi-annual
\xmm\ observations in 2017 and 2018 that supplement our annual
\chandra\ monitoring. A log of the observations obtained since 2016 is
presented in Table~\ref{obslog}.  Here we describe these data sets.

We concentrate on \xmm\ data obtained with the European Photon Imaging
Camera (EPIC) pn and MOS detectors.  Data from the Reflection Grating
Spectrometers are not used in this work.  The EPIC~pn
\citep{Struder01} sits at the focal plane of a coaligned, multi-nested
foil mirror with an on-axis point spread function with FWHM of
$\approx12\farcs5$ at 1.5~keV.  The EPIC instruments are sensitive to
X-rays in the 0.15$-$12~keV range with moderate energy resolution of
$E/\Delta E({\rm pn}) \sim $20$-$50.  In order to resolve the 0.424~s
pulse of \cco, the EPIC~pn data were obtained in {\tt
  PrimeSmallWindow} mode ($4\farcm3\times4\farcm3$), which has high
time resolution of 6~ms at the expense of 29\% deadtime.  Data
acquired with the two MOS detectors \citep{Turner01} were obtained in
{\tt PrimePartialW2} small-window mode on the central CCD with a
$1\farcm8\times1\farcm8$ FoV. The time resolution in this mode is
0.3~s, insufficient to resolve the pulsations of \cco. The MOS
cameras, less sensitive at the lower energy range of \cco, were used
only to confirm the EPIC~pn spectral results.

The \xmm\ data were reduced and analyzed using the Standard Analysis
Software (SAS) version 15.0.0 with the most up-to-date calibration
files. After filtering out background flares we obtained usable
exposure times listed in Table~\ref{obslog}. For the timing analysis
all photon arrival times were converted to barycentric dynamical time
(TDB) using the DE405 solar system ephemeris and the \chandra\
coordinates in \citet{got13a}.

We also examined the \chandra\ observations  that fell within the
post-glitch time interval (see Table~\ref{obslog}).  The pulsar was
placed on the S3 CCD of the Advanced Camera for Imaging and
Spectroscopy (ACIS), which was run in continuous-clocking mode that provides a
time resolution of 2.85~ms. We processed this data set following the
method outlined in \cite{got07} and \cite{hal11}.


\begin{deluxetable}{llcrc}
\tablewidth{0pt}
\tablecolumns{5}
\tablecaption{Log of New X-ray Timing Observations of \cco}
\tablehead{
\colhead{Mission}  & \colhead{Instrument}   & \colhead{ObsID}   & \colhead{Date} & \colhead{Exposure} \\
\colhead{} & \colhead{/Mode} & \colhead{} & \colhead{(UT)} & \colhead{(ks)} 
}
\startdata
{\it XMM\/} & EPIC-pn/SW &0780000201 &  2016\phantom{ii}  Jul 28  &  34.0  \\  
{\it XMM\/} & EPIC-pn/SW &0800960201 &  2017\phantom{i}  Jun 22   &  34.8  \\
{\it XMM\/} & EPIC-pn/SW &0800960301 &  2017\phantom{i}  Jun 23   &  22.2  \\ 
{\it XMM\/} & EPIC-pn/SW &0800960401 &  2017\phantom{i}  Jun 24   &  24.1  \\ 
{\it XMM\/} & EPIC-pn/SW &0800960501 &  2017\phantom{ii}  Jul 03  &  25.0  \\   
{\it XMM\/} & EPIC-pn/SW &0800960601 &  2017  Aug 10              &  21.3  \\   
\chandra\   & ACIS-S3/CC &     19612 &  2017\phantom{i}  Oct 10   &  33.0  \\
{\it XMM\/} & EPIC-pn/SW &0800960701 &  2017\phantom{i}  Dec 24   &  21.3  \\ 
{\it XMM\/} & EPIC-pn/SW &0821940201 &  2018\phantom{i}  Jun 22   &  33.4  \\
\chandra\   & ACIS-S3/CC &     19613 &  2018 Aug 27               &  \phantom{0}66.6   
\enddata
\label{obslog}
\end{deluxetable}


\begin{figure}
\centerline{\psfig{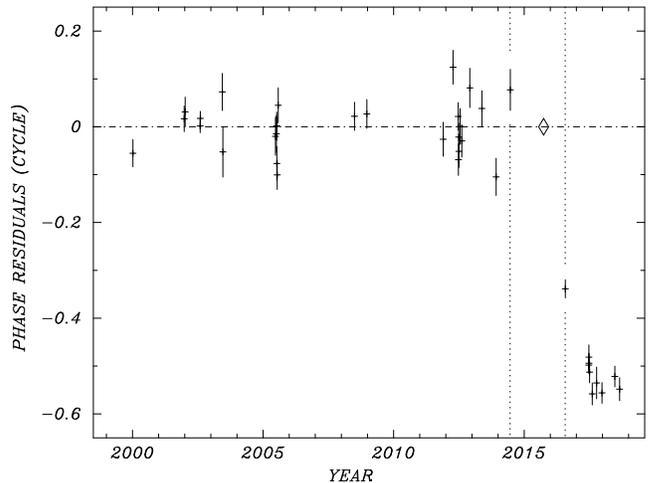}}
\caption{ \footnotesize Pulse-phase residuals from the timing solution
  of \citet{hal15}, with the addition of the ten post-2015 \xmm\ data
  points. A glitch occurred some time during the interval denoted by
  the vertical dotted lines. By matching the phase of the pre- and
  post-glitch ephemerides, the estimated glitch time indicated by the
  diamond is 2015 September~30.}
\label{fig:postres}
\end{figure}

\section{Timing Analysis\label{sec:timing}}

Prior to 2015, a unique, quadratic ephemeris from 
\xmm\ and \chandra\ observations of \cco\
adequately described its rotation
for 14~years \citep{hal15}.  Then, on 2016 July 28, we 
found a large ($\Delta\phi\approx-0.35$) and significant (18$\sigma$)
deviation in phase between the observed pulse arrival time and that
predicted from the pre-2015 ephemeris, consistent with the pulse 
arriving earlier than expected.  Using the subsequent observations,
we generated additional pulse times-of-arrival (TOAs)
following the recipe given in our previous papers
\citep{hal11,got13a} and continued to compare them to the pre-2015
ephemeris.  The persistent deviation in phase over time is
clearly evident in Figure~\ref{fig:postres} as a
highly significant shift of up to $\approx-0.55$~cycles
from the pre-2015 ephemeris.  This is standard behavior for a glitch,
namely, a speed-up in the spin frequency $f$ causes the pulses to come
successively earlier than the pre-glitch ephemeris predicts.

We also checked for any change in pulse shape and pulsed fraction,
in case what we think is a glitch might be due instead to a
change in the location of a hot spot on the surface, or the emergence
of a new emission process.
There is no significant difference between pre- and post-glitch
pulse shape, pulsed fraction, or energy dependence of pulse phase,
indicating that the surface thermal emission pattern has not changed.
So a simple glitch is the most straightforward interpretation of
Figure~\ref{fig:postres}.

Because of the long, 2-year gap in observations between 2014 and 2016,
we cannot be certain of the exact residual cycle count after the glitch,
nor determine the epoch of the glitch with any precision.
Fitting possible linear slopes to a few points after the glitch
in Figure~\ref{fig:postres} implies that the glitch magnitude is
$\Delta f/f = (2.8\pm0.4)\times10^{-9}$, which is the minimum possible
magnitude corresponding to the minimum possible phase shift, as plotted.
We also have independent evidence for an increase in frequency
by fitting a new, phase-connected quadratic ephemeris
to the ten post-glitch points from 2016 July~28 to 2018 Aug~27.
In this fit, $\Delta f/f_{\rm pred} = (5.22\pm0.80)\times10^{-9}$
(at a glitch epoch of MJD~57295 = 2015 September~30), which
is not precise enough to confirm the phase counting in
Figure~\ref{fig:postres}, but does suggests a larger glitch magnitude
than the simple linear fit.  Both pre- and post-glitch
ephemerides are listed in Table~\ref{tab:ephem}.

Post-glitch behavior typically includes partial recovery toward the
pre-glitch ephemeris, which can be fitted as a glitch also in $\dot f$
that subsequently decays on one or more time scales
(see examples in \citealt{esp11}).
In this context, our post-glitch ephemeris only crudely estimates
an average value for the post-glitch $\dot f$ because the data
points are not precise or numerous enough to track any change
in $\dot f$.  Nevertheless, a post-glitch $\dot f$ is significantly
detected at $(-2.82\pm0.31)\times10^{-16}$~s$^{-2}$, which is larger
than the historical value of $(-1.2398\pm0.0083)\times10^{-16}$~s$^{-2}$.

\begin{deluxetable}{lc}
\tighten
\tablewidth{0pt}
\tablecolumns{2}
\tablecaption{Ephemerides for \cco}
\tablehead{
\colhead{Parameter} & \colhead{Value$^{\rm a}$}
}
\startdata
R.A. (J2000)                                  & $12^{\rm h}10^{\rm m}00^{\rm s}\!.91$ \\
Decl. (J2000)                                 & $-52^{\circ}26^{\prime}28^{\prime\prime}\!.4$ \\
Surface dipole dipole field, $B_s$            & $9.8 \times 10^{10}$ G\\
Spin-down luminosity, $\dot E$                & $1.2 \times 10^{31}$ erg s$^{-1}$ \\
Characteristic age, $\tau_c\equiv P/2\dot P$  & 301 Myr \\
\cutinhead{Pre-glitch Timing Solution (2000-2014) \hfill}
Epoch of ephemeris (MJD TDB)                  & 53562.00000052 \\
Span of ephemeris (MJD)                       & 51549--56829 \\
Frequency, $f$                                & 2.357763502866(65)~s$^{-1}$ \\
Frequency derivative, $\dot f$                & $-1.2398(83) \times 10^{-16}$ s$^{-2}$ \\
Period, $P$                                   & 0.424130748815(12)~s \\
Period derivative, $\dot P$                   & $2.230(14) \times 10^{-17}$ \\
$\chi^2_{\nu}({\rm DoF})$                     &  2.68(25) \\
\cutinhead{Post-glitch Timing Solution (2016-2018) \hfill} 
Epoch of ephemeris (MJD TDB)                  & 57977.0000040 \\
Span of ephemeris (MJD)                       & 57597--58358 \\
Frequency, $f$                                & $2.35776345859(38)$~s$^{-1}$ \\
Frequency derivative, $\dot f$                & $-2.82(31) \times 10^{-16}$ s$^{-2}$ \\
Period, $P$                                   & $0.424130756780(68)$~s \\
Period derivative, $\dot P$                   & $5.06(56) \times 10^{-17}$ \\
$\chi^2_{\nu}({\rm DoF})$                     &  0.86(7) \\
\cutinhead{Glitch Parameters \hfill}
Epoch (MJD)                                   & 57295$^{\rm b}$ \\
$\Delta f$                                    & $(1.23\pm0.19)\times 10^{-8}$~s$^{-1}$ \\
$\Delta f/f_{\rm pred}$                       & $(5.22\pm0.80)\times10^{-9}$ \\
$\Delta \dot f$                               & $(-1.58\pm0.31)\times10^{-16}$ s$^{-2}$ \\
$\Delta \dot f/\dot f_{\rm pred}$             & $1.27\pm0.25$
\enddata
\tablenotetext{a}{Uncertainties in the last digits are given in parentheses.}
\tablenotetext{b}{Epoch for the glitch estimated by matching the zero phase 
of the two timing solutions; this assumes a constant post-glitch $\dot f$.}
\label{tab:ephem}
\end{deluxetable}

%

\section{Spectral Analysis\label{sec:spectra}}

We also examined the pre- and post-glitch \xmm\ spectra of \cco\ to
look for any change by comparing six observations obtained in 2012 and six in 2017.
For each EPIC~pn  observation we extracted spectra using an aperture of
radius $0\farcm5$ and a nearby off-source circular region of radius $1\farcm0$.
Response matrices and effective area files were generated for each
observation using the SAS software suite. We combined the spectra
extracted from the pre- and post-glitch observations, respectively,
using the FTOOL {\it addascaspec} to produce a single source spectrum
and associated files at each epoch.  These spectra were grouped to
include at least 200 counts per channel and were fitted using {\tt
  XSPEC} v12.10.0c software \citep{arn96}.  The two spectra were
fitted simultaneously to a two-blackbody model with interstellar
absorption and cyclotron lines in the range 0.3$-$2.5~keV.  We
characterize the column density using the {\tt TBabs} absorption
model, selecting the {\it wilm} Solar abundances \citep{wil00} and the
{\it vern} photoionization cross-section \citep{ver96}.

We note that it is not possible to uniquely fit for the column
density, softer blackbody $kT_1$, and the absorption features
simultaneously.  Furthermore, the detailed shape of the absorption
features is not known. We used a Gaussian line whose width is fixed
to the minimum value needed to characterize the absorption and then
fixed the column density to the average overall value for this fixed
Gaussian width. We consider this a representative model in order to
effectively compare the pre- and post-glitch spectrum.

\begin{figure}[t]
\centerline{\psfig{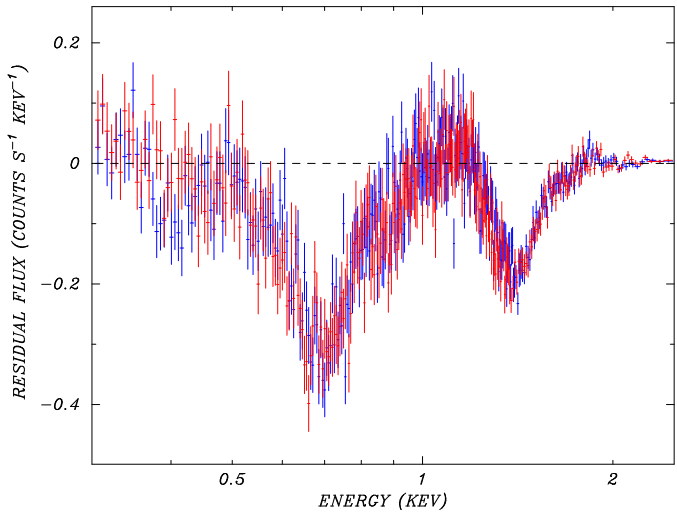}}
\caption{ \footnotesize Comparison of the pre- and post-glitch
  spectra of \cco\ using combined \xmm\ EPIC~pn data acquired in
  2012 (blue) and in 2017 (red). Residuals from the best fit pre-glitch 
  two-temperature blackbody model given in Table~\ref{tab:spectra}
  with the absorption lines removed are compared to the post-glitch
  spectrum using the same model parameters.
  Electron cyclotron features at 0.7~keV and 1.4~keV are evident, but
  there is no significant change in the line centroids, and thus the
  magnetic field strength, following the glitch.}
\label{fig:spec} 
\end{figure}

The results of the fits are presented in Table~\ref{tab:spectra}.  The
residuals from the pre-glitch two-blackbody model are shown in
Figure~\ref{fig:spec} to highlight any differences.  The prominent
absorption features are consistent with an electron cyclotron
fundamental ($E_0 = 0.7$~keV) and its harmonics in magnetic field of
$B \approx 8 \times 10^{10}$~G, according to the relation $E_0 =
1.16\,(B/10^{11}\,{\rm G})/(1+z)$~keV, where $z\approx0.3$ is the
gravitational redshift.  Within the statistics of the spectra, we find
no change in the spectrum from 2012 to 2017, specifically, no definite
shift in $E_0$ that would indicate a change in surface magnetic field
strength after the glitch. Instead, the field strength is seen to be
constant to $\approx 2\%$.  The two temperatures and total flux are
also unchanged, the latter at the $0.5\%$ level.

\begin{deluxetable}{lcc}
\tablewidth{0.0pt}
\tablecolumns{3}
\tablecaption{\cco\ \xmm\ Spectral Fitting}
\tablehead{
\colhead{Parameter} & \colhead{Pre-glitch\tablenotemark{a}} & \colhead{Post-glitch\tablenotemark{b}}
}
\startdata
Epoch                             & 2012                          &        2017                  \\
$N_{\rm H}$ (cm$^{-2}$) (fixed)   & $1.66\times 10^{21}$          & $1.66\times 10^{21}$         \\
$kT_1$ (keV)                      & $0.0801^{+0.0046}_{-0.0042}$  &  $0.0742^{+0.0042}_{-0.0037}$\\
$kT_2$ (keV)                      & $0.2513^{+0.0022}_{-0.0021}$  &  $0.2509^{+0.0021}_{-0.0021}$\\
$E_0$ (keV)                       & $0.712^{+0.012}_{-0.011}$     &  $0.710^{+0.017}_{-0.015}$\\
$\sigma_0$ (keV) (fixed)          & 0.08                          &   0.08\\
$\tau_0$                          & 0.26                          &   0.22\\
$E_1$ (keV)                       & $1.4292^{+0.0087}_{-0.0088}$  &   $1.4216^{+0.0089}_{-0.0090}$\\
$\sigma_1$  (keV) (fixed)         & 0.08                          &   0.08\\
$\tau_1$                          & 0.098                         &   0.10\\
$F_x ({\rm pn)}$\tablenotemark{c} & $2.084^{+0.010}_{-0.011}$     &  $2.078^{+0.010}_{-0.012}$\\
$\chi^2_{\nu}$(DoF)               & $1.39(283)$                   &   $1.32(276)$\\
\enddata
\tablenotetext{}{Notes --- EPIC~pn spectral fit in the 0.3$-$2.5~keV
  range, with the column density and Gaussian line widths held fixed
  (see text). Uncertainties on $kT$ and $E$ are given at the 68\%
  confidence level for 4 interesting parameters. Uncertainty on the
  flux is given at the 90\% confidence level.}
\tablenotetext{a}{Pre-glitch live time 107~ks from ObsIDs
  0679590101/201/301/401/501/601.}  \tablenotetext{b}{Post-glitch live
  time 98~ks from ObsIDs 0800960201/301/401/501/601/701.}
\tablenotetext{c}{Absorbed 0.3$-$2.5~keV flux in units of $10^{-12}$
  erg~cm$^{-2}$~s$^{-1}$.}
\label{tab:spectra}
\end{deluxetable}

\section{Discussion\label{sec:discussion}}

\subsection{Glitch Magnitude}

The distribution of glitch magnitudes in pulsars is bimodal,
with a broad peak centered at $\Delta f/f\approx3\times10^{-9}$,
and a narrow peak at $\Delta f/f\approx1\times10^{-6}$ \citep{esp11}.
The glitch in \cco\ is thus typical of the lower-amplitude group,
and also of glitch sizes in the Crab pulsar.  However, it is
unprecedented for a pulsar with the timing properties of
\cco\ to even have a glitch.  Glitch activity
correlates best with the spin-down rate of pulsars
in a linear manner such that
the average amount of spin-down reversed in a glitch, $\dot f_g$,
is $\approx0.01|\dot f|$, i.e., $1\%$ of the long-term spin-down
is reversed \citep{esp11,fue18}.  This has been interpreted 
in terms of the the vortex creep theory \citep{alp84} to
imply that $1\%$ or more of the moment of inertia of the
NS is contained in a crustal superfluid whose vortices are
repeatedly pinned and unpinned.

In this picture, it is natural that pulsars with small
$\dot f$ would not glitch frequently.
Among pulsars with $|\dot f|<1\times10^{-15}$~s$^{-2}$,
only three glitches have been observed in $\approx4260$
pulsar-years of monitoring, and no pulsar with
$|\dot f|<3\times10^{-16}$~s$^{-2}$ has glitched in $1780$~yr
\citep{fue18}.  
For \cco\ with its $\dot f=-1.24\times10^{-16}$~s$^{-2}$
to have glitched once in 15 years of monitoring makes it,
therefore, a significant outlier with $99\%$ confidence.

In addition, typical glitches in $\Delta \dot f/\dot f$ are only
$\sim 10^{-3}-10^{-2}$, much smaller than the order unity 
change in \cco\ if we take its post-glitch ephemeris literally.
In magnetars, however, it is common to see glitches in $\dot f$
of order unity \citep{kas03,dib08,dib14}, but these
could be due to magnetospheric phenomena such as
particle winds and rearrangement of twisted magnetic
field lines that change the dipole moment by a large
factor.  In CCOs there is no magnetospheric activity
and, uniquely testable in the case of \cco, 
no change in the surface dipole magnetic field strength
from the energies of the cyclotron lines.
Therefore, it is more likely that the change in
$\dot f$, if real, is due to internal torques between
the superfluid and normal matter.  However, $\Delta\dot f/\dot f$
is too large to be explained by vortex pinning
unless most of the moment of inertia of the NS can
become pinned.

{ There is a class of intermittent radio pulsars, whose $|\dot f|$
is larger by a factor of order unity when they are turned
on as radio pulsars in comparison with their off states
(e.g., \citealt{lyn17}).  States can persist for days up to years,
with the most extreme example being that of PSR J1841$-$0500
\citep{cam12}, which was off for 540 days.  Its $|\dot f|$ was
higher by a factor of 2.5 while the radio pulsations were turned on.
The implication is that a magnetospheric plasma is present and
contributing to the spin-down torque only during the radio on state.
If the glitch triggered a transition to a magnetospherically
active state in \cco, then its higher $|\dot f|$ could be
an indicator that it has turned on as a radio pulsar.}
We reserve judgment on this unexpected result until future observations 
can measure a new, long-term value of $\dot f$ with a precision comparable 
to the pre-glitch ephemeris.

\subsection{CCO Structure and Evolution}

Although the glitch activity of \cco\ would not have been
predicted from its small $\dot f$
and weak surface dipole field $B_s$, its internal magnetic field
may be as strong as those of canonical young pulsars that do glitch,
according to two arguments.
First, the thermal X-ray pulsations by which CCO pulsars are discovered
are difficult to explain in the context of weak magnetic fields,
since the only mechanism thought to be capable of creating a
non-uniform surface temperature is anisotropic heat conduction in a
strong magnetic field.  The effects of different magnetic field
configurations on heat transport in the crust and envelope of NSs were
modeled by \citet{gep04}, \citet{gep06}, \citet{per06}, and \citet{pon09}.
A toroidal field is expected to be the initial configuration generated
by differential rotation in the proto-NS dynamo \citep{tho93}.
One of the effects of crustal toroidal field is to insulate the magnetic
equator from heat conduction, resulting in warm spots at the poles.
To have a significant effect on heat transport, the crustal toroidal
field strength required in all models is $>10^{14}$~G, many orders of
magnitude greater than the poloidal field if the latter is measured by
the spin-down. \cite{sha12} tried to model the pulse profile of the CCO
\kpsr\ in \ksnr\ with anisotropic conduction in such a model, and concluded
that they needed a toroidal crustal field of $B_{\phi} > 2 \times 10^{14}$~G
to achieve the large observed pulsed fraction of $64\pm2\%$, although
they could not actually match the broad pulse shape
(see also \citealt{bog14a}.) 

Second, a theory of CCOs posits that they are born with a canonical NS
magnetic field that was buried by fall-back of a small amount of
supernova ejecta, $\sim 10^{-4}\,M_{\odot}$, during the hours and days
after the explosion. The buried field will diffuse back to the surface
on a timescale of $\sim10^5$~yr
\citep{gep99,ber10,ber13,ho11,ho15,vig12}.  During this time the CCO
will move vertically up in the $P-\dot P$ diagram due to its rapidly
increased braking as the dipole field grows \citep[see Figure~1
of][]{luo15}.  Eventually it will join the bulk of the population of
ordinary radio pulsars.  Such a scenario addresses the absence of
CCO descendants that should remain in the same region of $P-\dot P$
space long after their natal SNRs fade, if their weak magnetic
fields are intrinsic.  Searches for a thermal X-ray signature 
from such CCO descendants have yet to find a single example
\citep{got13b,bog14b,luo15}, except {\it possibly} for the unusual X-ray
pulsar Calvera \citep{hal13,hal15}.  The field growth hypothesis also
has the feature of not requiring yet another class of NS to exist that
would only exacerbate the apparent excess of pulsars with respect to
the Galactic core-collapse supernova rate \citep{kea08}.

More generally, magnetic field growth { \citep{bla83} or
dipole axis counteralignment \citep{mac74} have long been considered
possible reasons} why most measured pulsar braking indices,
defined as $n\equiv f\ddot f/\dot f^2$, are less
than the static dipole value of 3.  If the dipole magnetic field strength
$B_s$ is increasing at the rate $\dot B_s$, the braking index is reduced
to $$n = 3 - 2\,{\dot B_s f \over B_s |\dot f|}.$$
\cite{ho15} calculated that the upward vectors in the $P-\dot P$
diagram of three pulsars with $n<1.7$, namely,
PSR B0833$-$45 (Vela), PSR J0537$-$6910 in the LMC, and PSR J1734$-$3333,
could be explained by the same processes of field burial and diffusion
as in CCOs, but with a smaller amount of accreted mass,
$\sim10^{-5}\,M_{\odot}$.  These three pulsars also happen to have
large glitches, in the upper peak of the bimodal distribution
of glitch magnitudes.
The possible connection of low braking
index to regular, large glitches, led \cite{ho15} to propose that
this type of glitch activity could be triggered by the motion of
magnetic fields through the NS crust, interacting with the neutron
superfluid there.  Although only a small glitch has been seen
from \cco, it is interesting to consider that,
even though there is no evidence that glitch activity is correlated
with {\it dipolar} magnetic field strength \citep{fue18},
a glitch may be triggered by motion of {\it internal\/} magnetic field.

\subsection{Accretion Torque Noise?}

Here we consider an alternative interpretation of the apparent
glitch: torque noise during low-level accretion, possibly from
a fall-back debris disk, following the arguments in \citet{hal07}.
While the $B$-field of \cco\ derived
from timing only assumes dipole braking, its spin parameters fall
in a regime where both dipole braking and accretion disk torques are
conceivably significant.  Accretion at a rate $\dot
M=10^{11}$~g~s$^{-1}$ (or less if the NS magnetic field is weaker) can
penetrate the light cylinder to the magnetospheric radius.  If so, the
system is in the propeller regime, in which matter flung out takes
angular momentum from the NS, causing it to spin down.  The propeller
spin-down rate is
$$\dot f \approx -9.5 \times 10^{-15}\,\mu_{29}^{8/7}\,\dot M_{13}^{3/7}
\left({M_{\rm NS} \over M_{\odot}}\right)^{-2/7}$$
$$\times\ I_{45}^{-1}\left({f\over2.357\ {\rm Hz}}\right)
\left(1- {f_{\rm eq} \over f}\right)\ {\rm s}^{-2}$$ \citep{men99}.  
{ Here $I_{45}$ is the NS moment of inertia in units
  of $10^{45}$ g~cm$^2$, $\mu_{29} = B_s\,R^3_{\rm NS}$ is the
  magnetic dipole moment in units of $10^{29}$ G~cm$^3$, $\dot M_{13}$
  is the mass transfer rate in units of $10^{13}$ g~s$^{-1}$, and
  $f_{\rm eq}$ is the equilibrium spin frequency, presumed to be $<<f$
  because of the young age and small $\dot M$.}  In this model $\dot
M$ is the rate of mass expelled, which must be $>\dot m$, the
accretion rate onto the NS. While $\dot m$ must be $<3\times10^{13}$
g~s$^{-1}$ so as not to exceed the bolometric luminosity of \cco,
$\approx 2.5\times10^{33}$~erg~s$^{-1}$, even such a small accretion
rate is ruled out by upper limits on accretion-disk luminosity from
the non-detection of an optical counterpart by {\it HST\/}
\citep{del11}.  The latter observation requires an even smaller $\dot
M<10^{12}$ g~s$^{-1}$.

Assuming that $\dot M\approx10^{11}$ g~s$^{-1}$, accretion contributes
negligibly to the luminosity, thus not violating the
upper limits on X-ray variability, while still allowing
$\dot f \approx -1.3 \times 10^{-15}$~s$^{-2}$ from the
propeller effect.  This is a factor of $\approx$4 greater than
the $\dot f$ of the post-glitch ephemeris.
Therefore, fluctuations in the propeller $\dot f$ cannot
be immediately ruled out as an explanation for the observed
timing irregularity.  Evolutionary models of fall-back disks
predict that an initial disk mass of $<10^{-6}\,M_{\odot}$
can easily supply, at the estimated 7000~yr present age of 
\pks, accretion at the above assumed rate \citep{del11}.

Whether the NS can acquire such a disk from the homologously expanding
ejecta moving with it, or from reverse-shocked SN ejecta,
depends on details of the explosion.  However, if $\sim10^{-4}\,M_{\odot}$
of debris can fall directly onto the NS to bury its magnetic field,
it is likely that $10^{-6}\,M_{\odot}$ can end up
in a disk because of its angular momentum.  Not so obvious is
the spectrum of timing noise produced at such a low
accretion rate, a regime which has thus far not been observed.
For now, the overall resemblance of the timing residuals
to a classic glitch profile 
in an isolated pulsar leads us to prefer the glitch model
over accretion torque noise.

\section{Conclusions and Future Work}

We have detected the first glitch in a CCO pulsar, \cco.  Its
frequency jump is at least $\Delta f/f = (2.8\pm0.4)\times10^{-9}$,
and possibly larger if a fitted change in $\dot f$ by a factor of 2.3
is real.  It is crucial to continue timing the pulsar to establish the
post-glitch $\dot f$ more accurately.  A radio pulsation search should
be made to test for new magnetospheric activity.  There is no evidence
for a change in the weak surface magnetic field from the X-ray
cyclotron features, and no change in luminosity.  Old pulsars with
$\dot f$ as small as that of \cco\ have not been seen to glitch, which
implies that the glitch mechanism is contingent upon an internal
property of this young pulsar, such as high magnetic field strength or
temperature.  \cite{ho15} suggested that diffusion of a previously
buried $B$-field could be the trigger for a glitch.  \cite{ho15}
further suggested that glitches could identify the missing descendants
of CCOs.  Finding descendants among the ordinary radio pulsar
population would solve a major observational problem in CCO evolution,
and provide strong support to the field-burial theory of their origin.

\acknowledgments

This investigation is based on observations obtained with \xmm , an
ESA science mission with instruments and contributions directly funded
by ESA Member States and NASA.  Support for this work was provided by
NASA through {\it XMM\/} grants NNX17AC12G and 80NSSC18K0452, and {\it
  Chandra} Award SAO GO7-18063X issued by the {\it Chandra} X-ray
Observatory Center, which is operated by the Smithsonian Astrophysical
Observatory for and on behalf of NASA under contract NAS8-03060.

\end{document}